\documentclass[sigconf]{acmart}
\AtBeginDocument{%
  \providecommand\BibTeX{{%
    Bib\TeX}}}

\copyrightyear{2026}
\acmYear{2026}
\setcopyright{cc}
\setcctype{by}
\acmConference[MSR '26]{23rd International Conference on Mining Software Repositories}{April 13--14, 2026}{Rio de Janeiro, Brazil}
\acmBooktitle{23rd International Conference on Mining Software Repositories (MSR '26), April 13--14, 2026, Rio de Janeiro, Brazil}
\acmPrice{}
\acmDOI{10.1145/3793302.3793339}
\acmISBN{979-8-4007-2474-9/2026/04}

\usepackage[ruled,lined,linesnumbered]{algorithm2e}
\usepackage{graphicx}
\usepackage{textcomp}
\usepackage{multirow}
\usepackage{url}
\usepackage{framed}
\usepackage{subcaption}
\usepackage{xcolor}

\definecolor{shadecolor}{RGB}{245,245,245}

\newcommand{\rvz}[1]{{#1}}

\newenvironment{shadedframe}{%
  \MakeFramed{\advance\hsize-\width\FrameRestore}}%
{\endMakeFramed}

\def\BibTeX{{\rm B\kern-.05em{\sc i\kern-.025em b}\kern-.08em
    T\kern-.1667em\lower.7ex\hbox{E}\kern-.125emX}}

\def\Underline{\setbox0\hbox\bgroup\let\\\endUnderline}
\def\endUnderline{\vphantom{y}\egroup\smash{\underline{\box0}}\\}
\def\|{\verb|}

\newcommand{\D}[2]{
  \mathrm{D}(\mathrm{MS}_{#1#2}, \mathrm{MS}_{#2#1})
}

\usepackage{xcolor}

\begin{document}

\title{Quantifying Competitive Relationships Among Open-Source Software Projects}

\author{Yuki Takei}
\orcid{0009-0000-3230-201X}
\email{s2540011@jaist.ac.jp}
\affiliation{%
  \institution{Japan Advanced Institute of Science and Technology}
  \city{Nomi}
  \state{Ishikawa}
  \country{Japan}
}

\author{Toshiaki Aoki}
\orcid{0000-0002-1209-6375}
\email{toshiaki@jaist.ac.jp}
\affiliation{%
  \institution{Japan Advanced Institute of Science and Technology}
  \city{Nomi}
  \state{Ishikawa}
  \country{Japan}
}

\author{Chaiyong Ragkhitwetsagul}
\orcid{0000-0002-6502-1107}
\email{chaiyong.rag@mahidol.ac.th}
\affiliation{%
  \institution{Faculty of ICT, Mahidol University}
  \city{Nakhon Pathom}
  \country{Thailand}
}

\renewcommand{\shortauthors}{Takei et al.}

\begin{abstract}
Throughout the history of software, evolution has occurred in cycles of rise and fall driven by competition, and open-source software (OSS) is no exception. This cycle is accelerating, particularly in rapidly evolving domains such as web development and deep learning. However, the impact of competitive relationships among OSS projects on their survival remains unclear, and there are risks of losing a competitive edge to rivals. To address this, this study proposes a new automated method called ``Mutual Impact Analysis of OSS (MIAO)'' to quantify these competitive relationships. The proposed method employs a structural vector autoregressive model and impulse response functions, normally used in macroeconomic analysis, to analyze the interactions among OSS projects. In an empirical analysis involving mining and analyzing 187 OSS project groups, MIAO identified projects that were forced to cease development owing to competitive influences with up to \rvz{78\%} accuracy, and the resulting features supported predictive experiments that anticipate cessation one year ahead with up to \rvz{74\%} accuracy. This suggests that MIAO could be a valuable tool for OSS project maintainers to understand the dynamics of OSS ecosystems and predict the rise and fall of OSS projects.
\end{abstract}

\ccsdesc[500]{Information systems~Data mining}
\ccsdesc[500]{Software and its engineering~Software evolution}
\ccsdesc[500]{Mathematics of computing~Statistical paradigms}

\keywords{Open Source Software, Competitive Analysis, Econometrics, Time-Series Analysis, SVAR Model, Software Project Survival}

\maketitle

\section{Introduction}
Open-source software (OSS) serves as a crucial digital infrastructure, and many information systems often comprise numerous OSS components at the operating system, language, and application levels. A report from Open UK \cite{openuk} states that 97\% of private companies and public sector organizations reported using OSS internally \cite{state-of-open-phase-two2021}. Similarly, a Black Duck audit \cite{BLACKDUCK2025} found that 97\% of 1,658 scanned projects contained open-source code, which accounted for 70\% of the total code.

Given that many modern information systems depend heavily on OSS, selecting an OSS project with high survivability is part of an effective strategy for the long-term maintenance and growth of such systems. However, questions persist regarding the sustainability of OSS projects, and many of them, including popular ones, did not last long \cite{Khondhu2013}. Throughout software history, emerging OSS projects have evolved by addressing the issues of existing OSS and proprietary ones while introducing new challenges.
For example, traditional virtualization (like VMware) improved server resource use but had issues with performance overhead and resource management. Containerization technologies, such as Docker, addressed these by offering lighter, faster application deployment. However, Docker then created new challenges, like the complexity of container image security and the need for large-scale orchestration (e.g., Kubernetes). 

This phenomenon illustrates a continuous cycle of technological innovation: a solution becomes a new standard, which then reveals different issues, prompting the next wave of technologies and approaches to overcome them.
In other words, OSS has its ups and downs, and even with a healthy community and project management or high code quality, there comes a time when it cannot withstand changing circumstances.

Many previous studies on OSS sustainability have predominantly focused on internal factors of the project. For example, Raja et al. \cite{6127835} indexed the viability of OSS projects based on three dimensions: vitality, resilience, and organization. Samoladas et al. \cite{SAMOLADAS2010902} conducted a survival analysis on the relationships among the activity level within OSS projects, developer dynamics, and project survival time. Liao et al. \cite{10.1007/s11036-018-0993-3} predicted project lifespans based on the influence of programming languages, number of files, and quality of core developers. However, the sustainability of a project is determined not only by internal factors but also by external factors. Coelho et al. \cite{Coelho_2017} identify ``obsolescence'' and the ``emergence of strong competitors'' as major external factors affecting the OSS projects' sustainability. 

\begin{figure}[t]
\centerline{\includegraphics[width=0.95\columnwidth]{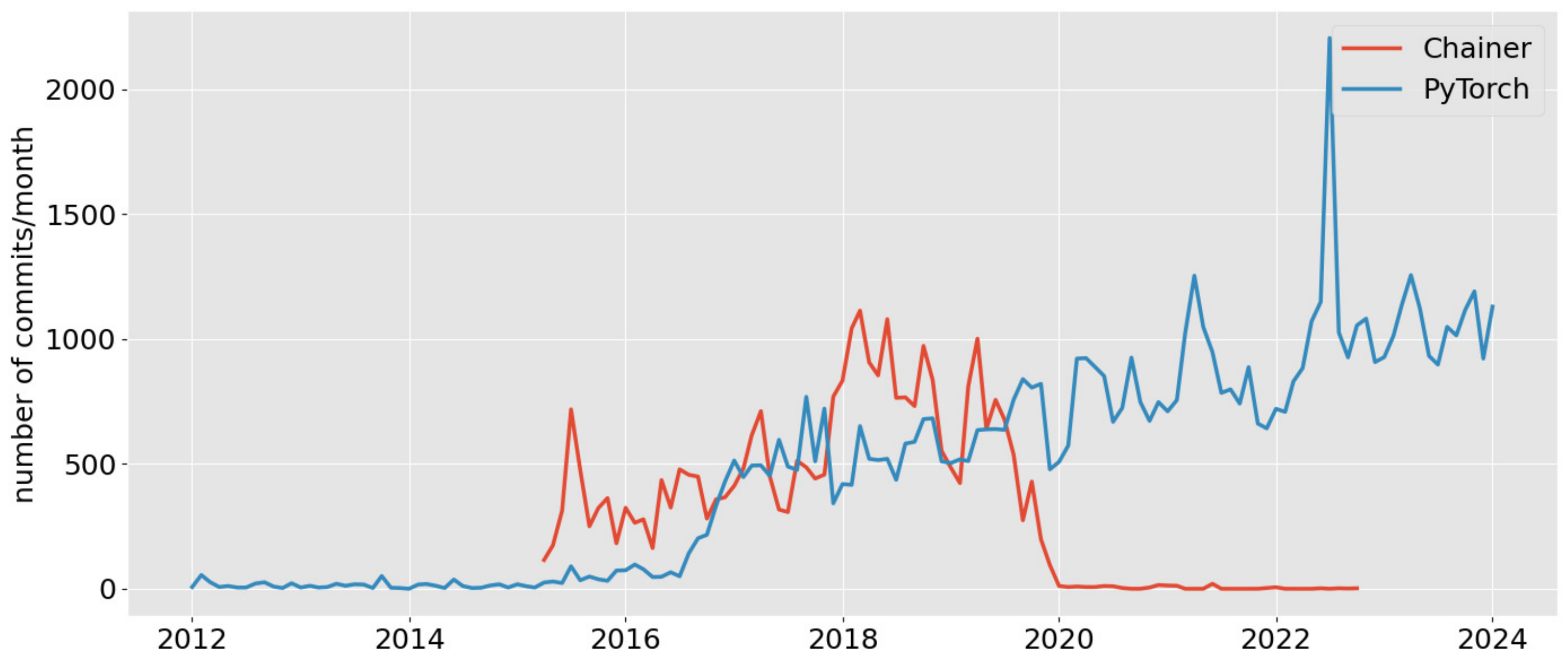}}
\caption{Growth of Chainer vs. PyTorch}
\vspace{-15pt}
\label{fig:chainer-vs-pytorch}
\end{figure}

The story of Chainer and PyTorch serves as a motivating example of OSS project competition and sustainability, driven by external factors. Chainer and PyTorch were once competing projects in the domain of deep learning frameworks. Later, Chainer eventually lost the development race to PyTorch and thus ceased new development~\cite{pr20191205}. The company behind Chainer made a statement that they \textit{``will stop new development of Chainer and move the entire company to PyTorch.''} In fact, from Fig. \ref{fig:chainer-vs-pytorch}, one can see how Chainer's activity drops sharply around 2019, while PyTorch's activity continues to rise. This example clearly shows that PyTorch is a strong competitor to Chainer, and its emergence made Chainer obsolete.

However, obsolescence and the emergence of strong competitors are uncertain events, and research on their impact on existing OSS projects remains limited. This is because quantifying OSS competitive relationships is challenging, as it requires complex causal inferences. 

To overcome these limitations, we propose a new method called ``\textbf{Mutual Impact Analysis of OSS (MIAO)}'' that \textit{quantifies the temporal competitive relationships among OSS projects using a structural vector autoregressive (SVAR) model and impulse response function (IRF).} By analyzing how changes in one project's activity affect others over time, MIAO enables us to identify and measure competitive dynamics that were previously difficult to quantify.

Accordingly, we formulate the following research questions and conduct a study to evaluate MIAO's ability to capture project survival and decline due to competitive pressure, aiming to provide actionable insights for OSS stakeholders.

\textbf{RQ1: How accurately can MIAO identify OSS projects that ceased development due to competitive influences?} 
By applying MIAO to a dataset comprising OSS projects that ceased development due to competition and those that did not, we can evaluate how well our method identifies and classifies these cases. In this context, we define a development cessation caused by competition as a ``Rising Event (REV)''. In other words, the classification task here is to distinguish between REV and non-REV cases.
This evaluation will help establish MIAO's practical utility for OSS stakeholders seeking to understand project sustainability.

\textbf{RQ2: What are the key factors of competitive relationships that affect the survival of OSS projects?} By examining the patterns and characteristics of competitive relationships in our dataset, we aim to identify which specific factors most strongly indicate that an OSS project is at risk of ceasing development due to competition.

This study makes the following contributions. First, we propose MIAO, a novel method to quantitatively measure temporal competitive relationships among open-source software projects. Second, we mined GitHub to create 187 OSS project groups focused on independent competitive relationships (i.e., excluding forks) and showed that MIAO reaches \rvz{78\%} accuracy in retrospective classification and \rvz{74\%} in one-year-ahead prediction. 
Finally, decision tree analysis revealed that unidirectional influence patterns characterize OSS decline.

\section{Theoretical background}

\subsection{Capturing Characteristics of OSS Data}

Competitive relationships among OSS projects emerge and evolve over time, so any analysis must employ methods that can accommodate temporal dynamics. Such methods must be able to formalize the interdependencies among multiple OSS projects and quantify their mutual impacts. 

The Structural Vector Autoregressive (SVAR) model~\cite{lutkepohl2005new}, a multivariate time‐series analysis technique typically used in macroeconomic analysis, is particularly well suited to this task.
Let us assume we have $n$ time-series datasets, each consisting of $m$ elements: $y_1 = [y_{1,t}, y_{1,t+1}, \dots, y_{1,t+(m-1)}], \hspace{5pt} \dots, \hspace{5pt} y_n = [y_{n,t}, y_{n,t+1}, \dots, y_{n,t+(m-1)}].$
Here, when we write $y_t$, it refers to the vector of $n$ elements at time $t$, that is, $y_t = (y_{1, t}, y_{2, t}, \dots, y_{n, t})'$. We consider expressing $y_t$ as a regression equation using the past $p$ lags of $y_t$, such as
\begin{equation}\label{svar-model}
B_0 y_t = c + B_1 y_{t-1} + B_2 y_{t-2} + \ldots + B_p y_{t-p} + u_t
\end{equation}

which we call an SVAR model. An SVAR model is derived by imposing structural constraints on a Vector Autoregressive (VAR) model. These structural constraints define how the variables instantaneously influence each other. In Equation~\eqref{svar-model}, $B_0$ is the matrix capturing this structure, $c$ is a constant vector, $B_i$ are coefficient matrices, and $u_t$ is an error term called the structural shock. From a stochastic process perspective, $u_t$ must be a vector of white noise.

If we interpret $y_t$ as time-series data representing the activity of $n$ OSS projects, then, for each OSS at time $t$, we can also model their activity using the past $p$ lags of activity. Optimal lag selection in VAR models is critical. Insufficient lags omit temporal dependencies, causing bias, while excessive lags increase parameters, risking overfitting. The ideal lag captures essential temporal dynamics while maintaining model parsimony for an optimal bias-variance balance.

\subsection{Formalizing Competitive Relationships Among OSS}

Our study mathematically formalizes such competitive relationships, which can be intuited from visual inspection of commit histories, through the use of Impulse Response Function (IRF)~\cite{lutkepohl2005new, hamilton1994series}. An IRF analyzes how a one-unit shock to the error term of a VAR-type model, as shown in Equation~\eqref{svar-model}, affects each variable (including itself and others) over time. Specifically, by converting a VAR or SVAR model into an infinite-order Vector Moving Average (VMA) process, the IRF emerges as the coefficients of that representation. The $\mathrm{VMA}(\infty)$ process~\cite{lutkepohl2005new, hamilton1994series} is expressed as: $y_t = \mu + B_0^{-1} u_t + \Psi_1 B_0^{-1} u_{t-1} + \Psi_2 B_0^{-1} u_{t-2} + \cdots$, where the coefficient matrices $\Psi_k B_0^{-1}$ applied to each structural shock $u_t$ correspond to the impulse responses. In other words, 
\begin{equation}
\label{orth-irf}
\text{IRF}_{ij}(k) = [\Psi_k B_0^{-1}]_{ij}
\end{equation}

represents the effect of a one-unit shock in variable $j$ at time $t$ on variable $i$ after $k$ periods. This makes it possible to quantitatively assess the dynamic interrelationships among different OSS over time.
Moqri et al.~\cite{Moqri02102018} demonstrated the value of IRF in open source community research by showing that gaining new followers causes a sustained positive effect on developer contributions that persists for approximately 10 months before gradually diminishing, effectively visualizing both the magnitude and temporal dynamics of social interaction effects.

Using the example of Chainer and PyTorch in Figure~\ref{fig:chainer-vs-pytorch}, suppose we represent their activities through an SVAR model and inject a one-unit shock (i.e., a sudden increase in the level of activity) into $u_t$ for Chainer and PyTorch, respectively. We can then mathematically evaluate the following. First, how PyTorch reacts to changes in Chainer’s activity. Second, how Chainer reacts to changes in PyTorch’s activity. Third, how those reactions evolve over time.

\section{Our Proposed Method: MIAO}

MIAO is \textit{a method for quantifying the impact on survival among OSS projects using IRF}. Because MIAO uses VAR and IRF, the variables representing OSS survival are expressed as time series data, referred to as \textit{activity time series data} in this study. Using this term, MIAO can be rephrased as capturing \textit{how shocks to one activity's time-series data propagate to other data over time as specific numerical values}. 

According to Kalliamvakou et al. \cite{10.1007/s10664-015-9393-5}, there is a positive correlation (0.61) between the number of active days and the number of commits in OSS projects, indicating that projects with more active days tend to have more commits. Therefore, considering this progression as activity time series data, MIAO quantifies the specific increase or decrease in the number of commits of other OSS projects over time following an increase in the commits of a particular OSS.

The MIAO algorithm comprises two main phases. Phase 1 involves calculating the IRF with multiple activity time series data as input. Phase 2 condenses the obtained IRF into MIAO scores, which are real values representing the relative magnitude of influence among OSS projects. We explain each phase below.

\subsection{MIAO Phase 1}
\label{sec:miao-phase1}

\begin{algorithm}[!t]
        \caption{MIAO Phase 1}
        \label{alg:miao-phase1}
        \KwIn{$\mathcal{T} = [T_1, T_2, \cdots, T_m], \mathcal{A} = [A_1, A_2, \cdots, A_n]$}
        \KwOut{$\mathcal{O} := 4\text{th order tensor of } m \times n \times n \times k$}
        
        $k := $ Period for calculating IRF ($k=1, 2, \cdots$)\;
        $max\_lag := $ Maximum lag order for VAR\;
        $ic := $ AIC or BIC or HQIC\;

        \SetKwProg{Fn}{function}{:}{end function}
        \Fn{\textnormal{PREPARE\_VAR}($\mathcal{D}$)\label{line:start:prepare-var}}{
            $\mathcal{D}' \leftarrow []$\;
            \ForEach{$A_{T_m} \in \mathcal{D}$\label{line:start:tm-loop}}{
                \If{$\mathrm{ADF\_TEST}(A_{T_m}) \geq 0.05$\label{line:start:adf-test}}{
                    $n \leftarrow$ Minimum number in $[0.1,1]$ where $\mathrm{FRAC\_DIFF}(A_{T_m}, n)$ becomes stationary\;
                    $A_{T_m} \leftarrow \mathrm{FRAC\_DIFF}(A_{T_m}, n)$\label{line:fracdiff}\;
                }\label{line:end:adf-test}
                $\mathcal{D'} \leftarrow \mathcal{D}' + [A_{T_m}]$\label{line:preprocess}\;
            }\label{line:end:tm-loop}

            $histories \leftarrow []$\label{line:histories-declare}\;
            $lags \leftarrow \mathrm{ICS}(\mathcal{D}', max\_lag, ic)$\label{line:ics}\;
            \For{$lag = 0$ \KwTo $|ics|$\label{line:start:ics-loop}}{
                $\_, \_, \_, u \leftarrow \mathrm{SVAR\_FIT}(\mathcal{D}', lag)$\label{line:svar-fit-prepare}\;
                $p \leftarrow \mathrm{WHITENESS\_TEST}(u)$\label{line:whiteness-test}\;
                $histories \leftarrow histories + [(ics[lag], lag, p)]$\;
            }\label{line:end:ics-loop}

            $lag \leftarrow \mathrm{VAR\_BEST\_LAG}(histories)$\label{line:var_best_lag}\;

            \Return $lag, \mathcal{D}'$\;
        }\label{line:end:prepare-var}

        \Fn{\textnormal{MIAO\_P1}($\mathcal{A}, T_m$)}{
            $\mathcal{D} \leftarrow []$\;
            
            \ForEach{$A \in \mathcal{A}$\label{line:start:a-roop}}{
                $A_{T_m} \leftarrow \mathrm{SUBSEQUENCE}(A, T_m)$\label{line:subseqence}\;
                $\mathcal{D} \leftarrow \mathcal{D} + [A_{T_m}]$\;
            }\label{line:end:a-roop}

            $lag, \mathcal{D}' \leftarrow \mathrm{PREPARE\_VAR}(\mathcal{D})$\label{line:prepare-var}\;
            $B_0, B, c, u \leftarrow \mathrm{SVAR\_FIT}(\mathcal{D}', lag)$\label{line:svar-fit2}\;

            \Return $\mathrm{IRF}(B_0, B, c, u, k, \mathcal{D}')$\label{line:irf}\;
        }

        \ForEach{$T_m \in \mathcal{T}$\label{line:start:miao}}{
            $\mathcal{O} \leftarrow \mathcal{O} + [ \mathrm{MIAO\_P1}(\mathcal{A}, T_m) ]$\;
        }\label{line:end:miao}
\end{algorithm}

Phase 1 follows the general IRF calculation process~\cite{lutkepohl2005new, hamilton1994series}. It includes the following steps: 1) confirming the stationarity of the time series, 2) verifying the presence of cointegration, 3) estimating the optimal lag order for VAR, 4) estimating VAR, and 5) calculating the IRF. In addition, MIAO tracks changes in influence among OSS projects over time by dividing the analysis into multiple periods and repeating the IRF calculation process for each period.

The reason for performing period division is that OSS has a growth model. Kuwata et al.~\cite{KUWATA20151004} discovered that many OSS projects follow a life cycle of Born, Childhood, Adolescence, Adulthood, and Obsolescence. In this model, activity is vigorous during Childhood and Adolescence. However, by the Obsolete stage, activity ceases, and maintenance is no longer continued. Referring back to Fig. \ref{fig:chainer-vs-pytorch}, we can see that the figure shows the monthly number of commits for Chainer\cite{Chainer} and PyTorch\cite{PyTorch}. Based on the model of Kuwata et al.~\cite{KUWATA20151004}, Chainer has passed through Born, Childhood, Adolescence, and Adulthood, and can be considered Obsolete at the end. PyTorch, on the other hand, can be considered to be in the Adolescence phase. Thus, OSS activity may undergo irreversible structural changes, and it is important to account for this growth model in time-series analysis of OSS. In this study, to address such structural changes in the time series, we divide the analysis period into multiple intervals, score the competitive relationships within each interval, and finally aggregate the scores across all periods.

MIAO takes two primary inputs: the set of analysis periods $\mathcal{T} = \{T_1, T_2, \dots, T_m\}$, and the set of activity time series data $\mathcal{A}$.

Regarding the first input, we denote the analysis period as $T_m (m = 1, 2, 3, \cdots)$. For example, if we analyze daily data over two years from January 1, 2016, to December 31, 2017, divided into one-year periods, $\mathcal{T}=[T_1, T_2]$, where $T_1$ and $T_2$ are defined as follows: $T_1=\{2016/01/01, 2016/01/02 \cdots, 2016/12/31\}, T_2=\{2016/01/01, 2016/01/02 \allowbreak \cdots, 2017/12/31\}$. MIAO does not specify detailed conditions for the number of intervals or the length of each $T_m$; thus, any can be constructed.

Regarding the second input, $\mathcal{A}$, for a time series to qualify as activity time series data, it must represent the survival of an OSS project and be comparable across multiple OSS projects. Examples of activity time series data include the changes in the number of commits and the evolution of download counts for packages such as PyPI~\cite{pypi} and NPM~\cite{npm}. Therefore, any time series meeting this condition is acceptable. Moreover, all activity time series data input to MIAO must include the minimum to maximum time of the $T_m$ input simultaneously. We denote the activity time series data obtained from a specific OSS as $A_i \in \mathcal{A}$.

The output of MIAO is $\mathcal{O}$, which comprises impulse response calculations from $A_i$ to $A_j$ across each $T_m$ for $k$ periods. Where $A_j$ represents the activity time series data of another OSS project $j$ within $\mathcal{A}$. When the VAR input contains $n$ variables, impulse responses can be calculated for $n \times n$ combinations. If the $n=2$, we have four combinations: $A_1 \to A_1$, $A_1 \to A_2$, $A_2 \to A_1$, and $A_2 \to A_2$. For each of these combinations, $\mathrm{IRF}_{ij}(k)$ can be calculated, resulting in a sequence of $k$ elements for each $A_i \to A_j$ impulse response. Since this is performed across multiple $T_m$ periods, the final output $\mathcal{O}$ from MIAO Phase 1 is a $m \times n \times n \times k$ tensor. 

Given these premises, the details of the algorithm in MIAO Phase 1 are shown in Algorithm~\ref{alg:miao-phase1}. Although our approach is grounded in standard VAR analysis, it is distinctive in that it uses $T_m$ as an explicit input to account for the OSS growth model, and employs fractional differencing to preserve long-term memory in the original series, which helps capture long-term competitive relationships.

\subsubsection{Preprocessing of time series data}

Lines \ref{line:start:tm-loop} to \ref{line:end:tm-loop} handle necessary conversions, such as seasonal adjustment and transformation to the stationary process to the time series $A_{T_m}$, and store the converted $A_{T_m}$ in $\mathcal{D}'$. Among these, the PREPROCESS function (Line \ref{line:preprocess}) applies preprocessing, such as seasonal adjustment, if necessary.

\subsubsection{Confirming the stationarity of the time series}

The stationarity of a time series was determined through unit root testing. The most commonly employed method is the Dickey–Fuller (DF) test proposed by Dickey et al. \cite{536b60b4-98b6-3231-a3fe-ba8f035840d4}. 
The Augmented Dickey–Fuller (ADF) test~\cite{10.1093/biomet/71.3.599} is recommended for VAR because it can address more complex time series structures. Line \ref{line:start:adf-test} employs the ADF test to determine if the time series $A_{T_m}$ is stationary. The significance level is set at 5\%, and if the p-value obtained by the ADF test exceeds 0.05, the time series $A_{T_m}$ is deemed to have a unit root process and is subsequently converted to a stationary process. The standard approach for converting a unit root process to a stationary process involves taking the first-order integer difference $y_t - y_{t-1}$. However, Marcos \cite{Marcos} highlights that integer differencing series can excessively remove information from any data series. Therefore, this study employed fractional differencing instead of integer differencing to minimize information loss from the original series. This process is executed in line \ref{line:fracdiff}. The fractional difference $n$ represents the minimum value that can render $A_{T_m}$ stationary. If the time series is originally stationary, no further action is taken.

\subsubsection{Verifying the presence of cointegration}

Cointegration refers to the case where two (or more) non-stationary time series share a stable long-run equilibrium, such that a particular linear combination of them is stationary. If cointegration is present, fitting a VAR on differenced series can discard this long-run relationship and lead to misspecified dynamics. MIAO necessitates all time series to be stationary (where the mean and autocovariance are time-invariant); it omits step 3, which checks for the presence of cointegration.

\subsubsection{Estimating the optimal lag order and VAR}

From line \ref{line:histories-declare} onwards in the PREPARE\_VAR function, estimating the optimal lag order for VAR begins. The variable $histories$ contains an array of tuples, each representing the information criterion value, lag order, and the p-value from the WHITENESS\_TEST function. In this study, we utilize the Ljung–Box test \cite{ad3868f4-d5d8-32fe-9700-07b98018b139} as the implementation of our WHITENESS\_TEST, which is a portmanteau test designed to check for the absence of autocorrelation in a time series. The null hypothesis $H_0$ of the Ljung–Box test posits that all autocorrelations from lags 1 to n are 0, while the alternative hypothesis $H_1$ states that at least one autocorrelation from lags 1 to m is not 0. The Ljung–Box test is necessary for the error term $u_t$ because the error term is assumed to be white noise. If this assumption is not met, the reliability of the VAR estimation results decreases.

Line \ref{line:ics} returns a sequence of information criterion values corresponding to lag orders up to $max\_lag$ based on the set $ic$ (information criterion). Lines \ref{line:start:ics-loop} to \ref{line:end:ics-loop} perform SVAR estimation for each lag, conduct the WHITENESS\_TEST on the error term, and store the test results in $histories$.

Finally, line \ref{line:var_best_lag} uses the VAR\_BEST\_LAG function to select a Pareto optimal lag order. It identifies the lag that minimizes the information criterion under the constraint that the model's residuals are free of serial correlation, as determined by the Ljung-Box test. The PREPARE\_VAR function concludes by returning this selected lag order and the previously created $\mathcal{D}'$.

\subsubsection{Calculating the IRF}

The MIAO\_P1 function handles all calculation processes and returns the IRF calculation results. It extracts the subsequence corresponding to the $T_m$ period from the input activity time series data $A$ using the SUBSEQUENCE function and stores it in $\mathcal{D}$ (Lines \ref{line:start:a-roop} to \ref{line:end:a-roop}). Once $\mathcal{D}$ is complete, the optimal lag order and $\mathcal{D}'$ with various conversions are obtained using the previously described PREPARE\_VAR function. Using these, SVAR fitting is performed in line \ref{line:svar-fit2} to estimate the parameters in Equation~\ref{svar-model}. Finally, IRF is calculated using the parameters obtained from SVAR\_FIT as input. The IRF function computes the impulse response using Equation~\ref{orth-irf} and returns the result as $n \times n \times k$ tensor.

Finally, MIAO\_P1 is executed for each $T_m$ and the output is stored in $\mathcal{O}$ (Lines \ref{line:start:miao} to \ref{line:end:miao}).

\subsection{MIAO Phase 2}

The output of MIAO is $n \times n$ impulse responses for each $T_m$, with each containing values for $k$ periods. These results cannot be utilized as is. Our focus is on the medium- to long-term influence among OSS, and we do not need to pay much attention to short-term fluctuations in impulse responses. Therefore, we used the cumulative impulse response, representing the cumulative impact as a feature. In this study, the value of the cumulative impulse response when $k$ approaches infinity was defined as Equation~\ref{eq:sce}.

\begin{equation}\label{eq:sce}
\mathrm{SCE}_{ij} = \sum_{k=1}^\infty \mathrm{IRF}_{ij}(k)
\end{equation}
where SCE represents the \textit{shock cumulative effect}. Note, IRFs derived from stationary VAR models are guaranteed to converge at infinity \cite{hamilton1994series}.

As SCE is calculated for each $T_m$, $m$ SCEs are obtained for each $A_j \to A_i$. We denote the $m$-th $\mathrm{SCE}_{ij}$ as $\mathrm{SCE}_{ij}^m$, and define the MIAO score ($\mathrm{MS}_{ij}$) as Equation~\ref{eq:miao-score}.

\begin{equation}\label{eq:miao-score}
\mathrm{MS}_{ij} = (-1) \cdot \sum_{k=1}^m \mathrm{SCE}_{ji}^k
\end{equation}

\noindent
We multiply by -1 to invert the sign so that the MIAO score increases as the negative impact becomes larger. 
When $\mathrm{MS}_{ij}$ is close to 0, it indicates that the positive and negative impacts oscillate, and the impact is canceled out in the long term. Note that in Equation~\ref{eq:miao-score} the subscripts $ij$ are reversed between the left-hand side and the right-hand side. This is to make the direction of influence in the right-hand-side MIAO score intuitively interpretable as from $i$ to $j$.

The number of $\mathrm{MS}_{ij}$ obtained in MIAO Phase 2 is $n \times n$, where $n = |\mathcal{A}|$. However, as $\mathrm{MS}_{ij} (i = j)$ from oneself to oneself is unnecessary, $n \times n - n$ $\mathrm{MS}_{ij}$ are finally obtained and output as a MIAO score table, as shown in Table \ref{tab:miao-score}.

\begin{table}[t]
    \centering
     \caption{General MIAO Score Table}
    \begin{tabular}{|c|c|} \hline
    \textbf{Direction of Influence} & \textbf{MIAO Score} \\ \hline
    $A_1 \to A_2$ & $\mathrm{MS}_{12}$ \\
    $\vdots$ & $\vdots$ \\
    $A_1 \to A_n$ & $\mathrm{MS}_{1n}$ \\
    $A_2 \to A_1$ & $\mathrm{MS}_{21}$ \\
    $\vdots$ & $\vdots$ \\
    $A_n \to A_{n - 1}$ & $\mathrm{MS}_{n, n - 1}$ \\
    \hline
    \end{tabular}
    \label{tab:miao-score}
\end{table}

\subsection{Interpretation of MIAO Score Table}\label{subsection:meaning-of-miao}

\begin{table*}[t]
    \centering
    \caption{Interpretation of MIAO Score Table}
    \begin{tabular}{|c|c|c|} \hline
        \textbf{ID} & \textbf{Case} & \textbf{Meaning} \\ \hline
        M1 & \begin{tabular}{c}
        $\mathrm{MS}_{ij} > 0 \land \mathrm{MS}_{ji} \leq 0$
        \end{tabular} & \begin{tabular}{c}
        $A_i$ has a negative impact on $A_j$, and $\D{i}{j}$ represents the magnitude of the impact.
        \end{tabular}\\ \hline

        M2 & \begin{tabular}{c}
        $\mathrm{MS}_{ij} \leq 0 \land \mathrm{MS}_{ji} > 0$
        \end{tabular} & \begin{tabular}{c}
        $A_j$ has a negative impact on $A_i$, and $\D{i}{j}$ represents the magnitude of the impact.
        \end{tabular}\\ \hline

        M3 & \begin{tabular}{c}
        $\mathrm{MS}_{ij} \geq 0 \land \mathrm{MS}_{ji} \geq 0$
        \end{tabular} & \begin{tabular}{c}
        $A_i$ and $A_j$ mutually have a negative impact, and $\D{i}{j}$ represents the magnitude of the impact.
        \end{tabular}\\ \hline

        M4 & \begin{tabular}{c}
        $\mathrm{MS}_{ij} \leq 0 \land \mathrm{MS}_{ji} \leq 0$
        \end{tabular} & \begin{tabular}{c}
        $A_i$ and $A_j$ mutually have a positive impact, and $\D{i}{j}$ represents the magnitude of the impact.
        \end{tabular}\\ \hline
     \end{tabular}
    \label{tab:interpretation-of-score}
\end{table*}

Table \ref{tab:interpretation-of-score}  shows the competitive dynamics between projects. We categorize the relationships into four distinct patterns based on the signs of the MIAO scores (denoted as Cases in the table).  The $\mathrm{D}$ in the table is a distance function where $\mathrm{D}(a, b) = |a - b|$.
Table \ref{tab:interpretation-of-score} alone does not provide sufficient information to read the score table; therefore, we will add several supplements. To prepare for this, we introduce competitive relationships into the score table. Specifically, let $T = A_i$ be the analysis target and $C_i = A_i (A_i \neq T)$ be the $i$-th competitor to $T$. When written as $\mathrm{MS}_{C_iT}$, it represents the MIAO score of $C_i \to T$.

First, we should suspect the occurrence of REV when $T = A_i, C_i = A_j$, Case M1 is satisfied, $\D{C_i}{T}$ is the maximum in the table, and $\D{C_i}{C_j}$ is relatively small. The absence of mutual influence is indicated when $\D{i}{j} \fallingdotseq 0$ in any of M1 to M4. However, even in cases where M1 is not satisfied, as in M3 or M4, if $\D{i}{j}$ is large in the table, it suggests that $\mathrm{MS}_{ij} \gg \mathrm{MS}_{ji}$ or $\mathrm{MS}_{ji} \gg \mathrm{MS}_{ij}$, implying possible REV.

\section{Experiment}
\label{sec:experiment}

\begin{figure*}[tb]
\centering
\centerline{\includegraphics[width=1.3\columnwidth]{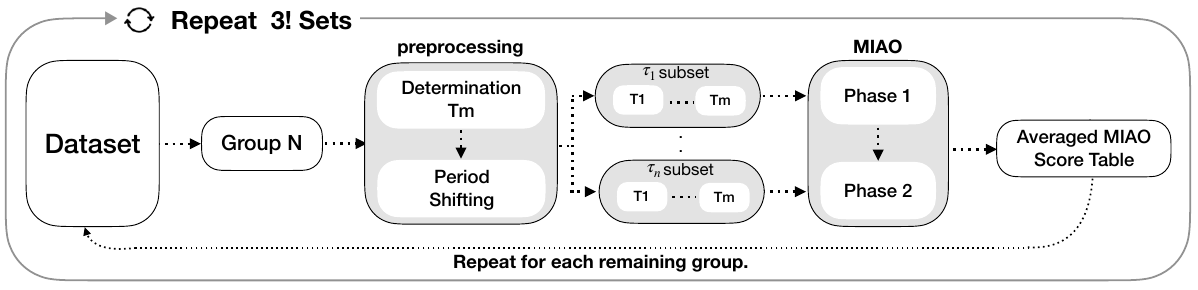}}
\caption{An overview of our experiments}
\label{fig:experiment-flow}
\end{figure*}

In this section, we present the results of a retrospective evaluation of MIAO using real‑world data. Figure~\ref{fig:experiment-flow} illustrates the overview of the experiment. We explain each step below.

\subsection{Dataset}

\begin{table*}[t]
    \centering
    \caption{An example of the Dataset}
    \resizebox{\textwidth}{!}{
        \begin{tabular}{|c|l|l|l|c|c|c|c|} \hline
        \textbf{Group} & \textbf{Target} & \textbf{Competitor1} & \textbf{Competitor2} & \textbf{REV} & \textbf{Start date} & \textbf{End date} & \textbf{\#split} \\ \hline
        1 & chainer/chainer & tensorflow/tensorflow & pytorch/pytorch & 1 & 2016-09-25 & 2020-09-25 & 1 \\
        & \vdots &  \vdots &  \vdots &  \vdots &  \vdots &  \vdots & \vdots \\
        187 & facebook/react-native & NativeScript/NativeScript & flutter/flutter & 0 & 2015-01-30 & 2023-01-30 & 2 \\
        \hline
        \end{tabular}
    }
    \label{tab:dataset}
\end{table*}

We constructed the full dataset by semi-automatically identifying known REV and non-REV GitHub projects (and their competitors), cloning their repositories, and extracting their commit logs. To reduce noise from automated activity, we excluded bot-generated commits by filtering out records whose author field contains the \texttt{[bot]} (e.g., \texttt{dependabot[bot]} and \texttt{github-actions[bot]}). Specifically, from the August 2024 data dump provided in the README of GitHub Search~\cite{ghs-dataset, Dabic2021}, we searched for projects with 2,538 or more stars, 23 or more contributors, and 500 or more commits. We designated projects that had ceased activity among the searched records as candidates for the REV class and those still active as candidates for the non-REV class. The search criteria of 2,538+ stars, 23+ contributors, and 500+ commits, high-threshold values derived from the statistics of the top 5,000 projects on GitHub, as determined by Coelho et al.\cite{Coelho_2017}, representing a group of OSS projects with activity levels that make them less likely to decline due to internal factors. Here, ``ceased activity'' is defined as satisfying one of the following conditions: 1) the repository is archived on GitHub, 2) there have been no commits for over a year, or 3) the average number of commits over the last 12 months is less than 1.5. The value of 1.5 in condition 3) is a slight relaxation of the criterion used by Valiev et al~\cite{10.1145/3236024.3236062}, adopted to increase the number of potential candidates for our analysis.

In our study, it is crucial to accurately identify which of these REV class candidates declined due to competitive pressure. Since explicit sources confirming cessation due to competition, like the announcement for Chainer, are rare, we needed to use other sources to identify competitors. For this study, we reviewed the README files of the REV class candidates and identified projects as belonging to the REV class if their README recommended migration to a competitor's OSS. For example, fizyr/keras-retinanet said \textit{``This repository is deprecated in favor of the torchvision module.''} \cite{keras-retinanet-deprecated}. This criterion yielded 234 OSS projects. 
Among the 234 REV candidates, the number of migration destinations (i.e., competing OSS projects) identifiable from README files and related materials was highly imbalanced. Specifically, 173 projects listed one competitor, 35 listed two, 16 listed three, and 10 listed four or more; the maximum was eleven (rustformers/llm). If we were to build VAR models whose dimensionality varies across cases, higher-dimensional cases would require estimating substantially more parameters, which tends to reduce estimation stability given finite time-series length. Moreover, comparing results across cases with different dimensionality undermines statistical comparability and makes interpretation inconsistent. Therefore, to ensure comparability across projects, we fixed the number of variables used in the analysis.

Fixing the dimensionality, however, involves an inherent trade-off. For example, if we adopted an 12-variable VAR (one target OSS plus eleven competitors) to accommodate cases with many identified competitors, then for projects with only one identified competitor we would need to additionally identify ten more competitors. Such forced augmentation would likely increase the risk of including projects that are not true competitors. Conversely, reducing all cases to one-to-one competition (one target plus one competitor) would discard how activity disperses across multiple alternatives in multi-competitor settings.

In our data, cases with three or more identified competitors account for only 11\% (26/234). Based on this distribution and the above trade-offs, we adopted a three-variable model consisting of one target OSS and two competitors, i.e., 1 target + 2 competitors. We constructed the two competitors per target as follows.
For the 35 projects with exactly two identified competitors, we used those two competitors as-is.
For the 26 projects with three or more identified competitors, we selected two competitors from the identified set by prioritizing those with the largest temporal overlap with the target OSS.
For the 173 projects with only one identified competitor, we supplemented one additional competitor to reach two competitors in total: we queried both ChatGPT o3 and Gemini 2.5 Pro for potential competitors, manually investigated projects that appeared in both responses, and added one competitor judged to be a plausible substitute within the same technical domain (e.g., TensorFlow for Chainer, or Ruby for Python). When multiple plausible candidates remained, we selected only the one with the largest temporal overlap with the target OSS.

We further excluded candidates that violated our analysis requirements or for which the decline was attributable to non-competitive factors, including:
(i) competitors not identifiable with sufficient confidence ($n{=}28$);
(ii) repository transfer/migration that directly explains the decline ($n{=}18$);
(iii) incorporation into an official or native replacement (e.g., absorbed into an official module or a non-public implementation, such as SwiftUI) ($n{=}10$);
(iv) temporal overlap with the identified competitors shorter than one year ($n{=}51$);
(v) inconsistencies with our activity dataset (e.g., fewer than 500 commits, or still alive status) ($n{=}3$); and (vi) fork-based competition ($n{=}39$). Among these, constraint (vi) warrants special attention.

In fork-based cases, the target and competitor share an identical pre-fork history; as a result, pre-fork observations exhibit perfect multicollinearity, preventing unique identification of VAR coefficients. We considered restricting the analysis to the post-fork period; however, in 29 out of the 39 fork-based cases, the parent project ceased maintenance immediately after the fork or the post-fork overlap period was shorter than one year, yielding insufficient observations under our minimum overlap requirement. Given these practical limitations, and to provide a robust baseline for this initial study, we intentionally scoped our analysis to \emph{independent} competitive relationships (e.g., Chainer vs.\ PyTorch) and excluded fork-based competition.

After applying these constraints, we ultimately selected 87 groups for the REV class, comprising 87 target projects and 174 competitor projects.

For the non-REV class, to avoid a significant imbalance with the sample size of the REV class, we selected the top 100 projects for which we successfully identified competitors, starting from the top of the non-REV pool sorted by star count. Since automatically identifying competitors for non-REV projects is challenging, we selected two appropriate competitors for each by combining queries to LLMs with manual investigation, as with the REV class, to form 3-variable groups. For these as well, we limited our selection to projects within the same technical domain and with a particularly large temporal overlap with the target. (e.g., Ruby on Rails, Laravel, Django)

\subsection{Determination of Analysis Period}\label{subsection:analysis-period}

The OSS in each dataset group is developed over different time scales, making a uniform analysis period impractical. Therefore, the analysis period must be determined individually for each group.

The start time of the analysis period for each group is the minimum of the dates that all activity time series data in the group commonly possess. Conversely, the determination of the end time differs between the REV and non-REV classes. First, for the REV class, the end time was set to the first date when the average number of commits over the most recent 12-month period, used to determine the cessation of activity, fell below 1.5. The end time is not set to the point of complete cessation of activity to enhance the sensitivity of the impulse response. For the non-REV class, the end time was set to the last date that the three variables have in common. Table~\ref{tab:dataset} shows examples of the start and end times determined for each group.

Next, we describe how to define $T_1, T_2, \cdots, T_m$ for each group. In this experiment, $T_m$ is created so that its elements increase by one year with each increment of $m$, starting from $m = 1$. In other words, at any point of $m$, the relationship $|T_m| \leq |T_{m+1}|$ holds. For example, if the start time is January 1, 2017, and the end time is December 31, 2018, $T_m$ would be constructed as $T_1 = \{2017/01/01, \cdots, 2017/12/31\}, T_2 = \{2017/01/01, \cdots, 2018/12/31\}$.

Notably, the longer the analysis period $T_m$, the more likely the time series is to have a trend. Especially when $m > 4$, serial correlation often appears in the error term of the VAR. Therefore, we limit the analysis period $T_m$ to a maximum of four years. If the analysis period exceeds this limit, we divide it into segments of four years and treat each segment as a different group. The number of these divisions is indicated by the column \#split in Table~\ref{tab:dataset}. For example, in Group 26, \#split is 6, meaning the period from the Start date to the End date is divided into 6 parts, and each is expanded as a different group of 6. 
Due to this period division, the number of groups for MIAO execution increased from 87 to 97 for the REV class, and from 100 to 279 for the non-REV class.

\subsection{Period Shift}\label{subsection:time-shift}

The impulse response is sensitive to fluctuations in the time series, and the results can change even if the analysis period shifts slightly back and forth. Therefore, to enhance the reliability and robustness of the experimental results, we constructed multiple MIAO models with different periods and used the average of the scores obtained from each MIAO model as the final score. This is a concept close to bagging in ensemble learning~\cite{Breiman1996Bagging}. Specifically, for the input $T_1, T_2, \cdots, T_m$, we shift the start time of each by $\tau_n$ days to construct a new analysis period and execute MIAO multiple times with these periods. We denote shifting $T_m$ by $\tau_n$ days as $T_{m \to \tau_n}$. For example, if we shift the start time of $T_m$ by one month, where $\tau_n = 1M$, we write it as $T_{m \rightarrow 1M}$. Therefore, $\mathrm{MS}_{ij}$ obtained from $T_{m \to \tau_n}$ can be expressed as $\mathrm{MS}_{ij \rightarrow \tau_n}$. If we set $T_{m \to 0}$, it is equivalent to the original $T_m$.

After n period shifts, we calculate the Average $\mathrm{MS}_{ij}$ ($\mathrm{AMS}_{ij}$)  as $\displaystyle \mathrm{AMS}_{ij} = \frac{\sum_{k=1}^{|\tau|} \mathrm{MS}_{ij \rightarrow \tau_k}}{|\tau|}$. In Figure~ \ref{fig:experiment-flow}, \textit{$\tau_i$ subset} represents the shifted dataset where columns $T_{m \to \tau_i}$ and corresponding $A_i$ serve as MIAO inputs.

\subsection{Parameter Settings}

The maximum $T_m$ is set to $T_4$ because larger values of m often introduce trends in the time series, violating the stationarity assumption required for VAR models. For period shifting, we apply four shifts: $T_{m \to 0}, T_{m \to 1M}, T_{m \to 2M}$ and $T_{m \to 3M}$, with a maximum shift of 3 months. Thus, for each $A_i \to A_j$ in each group, we obtain $\mathrm{AMS}_{ij}$, which is the average of $\mathrm{MS}_{ij \to 0}, \cdots, \mathrm{MS}_{ij \to 3M}$ for each $T_m$. To estimate the optimal lag order for VAR, we use the Akaike information criterion (AIC) \cite{akaike1973information}, selecting the optimal lag from 1 to 15 that minimizes the AIC. AIC provides a practical balance between goodness-of-fit and model complexity for noisy OSS data, making it well-suited for selecting a lag order that captures temporal dependence without excessive overfitting. However, if the residuals estimated at that lag order exhibit serial correlation, we select the smallest AIC lag without serial correlation. For the Ljung–Box test, if the lag estimated by AIC is less than 10, we use 10; if it is 10 or more, we use twice that lag order for testing, with a significance level of 10\%.

To capture contemporaneous competitor effects, we impose a recursive SVAR identification. A key implication of this recursive constraint is that the resulting impulse response functions depend critically on the ordering of the variables. In recursive SVAR identification, the ordering of variables determines the assumed causal direction. Accordingly, for the REV class, we fix the ordering as $\text{Competitor 1}\rightarrow\text{Competitor 2}\rightarrow\text{Target}$. So the target can receive contemporaneous influence from both competitors. The ordering determines the structure of \(B_{0}\) in Equation \eqref{orth-irf}, representing contemporaneous effects that are then propagated across the \(k\)-period impulse responses. In contrast, for the non-REV class, the true ordering is unknown; therefore, we perform a sensitivity analysis by evaluating all \(3!\) permutations of the three variables. Specifically, from the dataset in Table \ref{tab:dataset}, we fix all cases with REV=1 and create versions in which only the REV=0 cases are permuted. We then apply MIAO to each of these datasets and report the aggregated results. For both classes, $\text{Competitor 1}$ always has more commits than $\text{Competitor 2}$.

\section{Results}

\begin{table*}[t]
    \centering
    \caption{ Statistical results of tests and model estimations (Average of 3! permutations) }
    \resizebox{\textwidth}{!}{
    \begin{tabular}{|c|rrr|rrrrr|rrr|}\hline
        & \multicolumn{3}{c|}{\textbf{ADF Test and Fractional difference}} & \multicolumn{5}{c|}{\textbf{VAR Estimation}} & \multicolumn{3}{c|}{\textbf{ Ljung–Box Test}} \\
        \cline{2-12}
         & Test statistic & P-value & N-Frac diff & Sample size & Lag & AIC & BIC & HQIC & Maximum lag order & Test statistic & P-value \\ \hline
        Count & 15756.00 & 15756.00 & 15756.00 & 5252.00 & 5252.00 & 5252.00 & 5252.00 & 5252.00 & 5252.00 & 5252.00 & 5252.00 \\
        Mean & -8.01 & 0.01 & 0.02 & 850.45 & 10.93 & 7.36 & 8.04 & 7.63 & 20.54 & 96.57 & 0.38 \\
        Std. Dev. & 6.35 & 0.07 & 0.07 & 397.80 & 3.48 & 4.35 & 4.39 & 4.36 & 8.34 & 57.22 & 0.35 \\
        Min & -58.75 & 0.00 & 0.00 & 365.00 & 1.00 & -7.31 & -6.95 & -7.17 & 10.00 & 2.06 & 0.00 \\
        10\% & -17.19 & 0.00 & 0.00 & 365.00 & 7.00 & 1.80 & 2.45 & 2.03 & 10.00 & 22.91 & 0.00 \\
        25\% & -9.23 & 0.00 & 0.00 & 366.00 & 7.83 & 4.36 & 5.03 & 4.62 & 10.00 & 38.09 & 0.04 \\
        50\% & -5.56 & 0.00 & 0.00 & 731.00 & 10.83 & 7.46 & 8.10 & 7.72 & 21.67 & 99.60 & 0.30 \\
        75\% & -4.09 & 0.00 & 0.00 & 1096.00 & 14.00 & 10.25 & 10.96 & 10.51 & 28.00 & 139.73 & 0.70 \\
        90\% & -3.16 & 0.02 & 0.00 & 1461.00 & 15.00 & 12.97 & 13.58 & 13.19 & 30.00 & 169.06 & 0.94 \\
        Max & 1.93 & 1.00 & 0.80 & 1461.00 & 15.00 & 20.29 & 21.05 & 20.58 & 30.00 & 319.62 & 1.00 \\        
        \hline
    \end{tabular}
    }
    \label{tab:results-of-model-fitting}
\end{table*}

\begin{table}[t]
    \centering
    \caption{Example of score tables obtained from MIAO}
    \label{tab:results-of-miao}
    \begin{minipage}{0.45\columnwidth}
        \centering
        \begin{tabular}{|c|c|} \hline
            \textbf{Group 1} & $\pmb{\mathrm{AMS}_{ij}}$ \\ \hline
            pytorch $\to$ tensorflow & -4.44 \\
            chainer $\to$ tensorflow & -2.57 \\
            tensorflow $\to$ pytorch & -0.60 \\
            chainer $\to$ pytorch & -0.65 \\
            tensorflow $\to$ chainer & -0.21 \\
            pytorch $\to$ chainer & 5.90 \\
            \hline
        \end{tabular}
    \end{minipage}%
    \hfill
    \begin{minipage}{0.45\columnwidth}
        \centering
        \begin{tabular}{|c|c|} \hline
            \textbf{Group 24} & $\pmb{\mathrm{AMS}_{ij}}$ \\ \hline
            prefect $\to$ airflow & -0.20 \\
            orchest $\to$ airflow & -0.07 \\
            airflow $\to$ prefect & -3.70 \\
            orchest $\to$ prefect & -0.71 \\
            airflow $\to$ orchest & -2.98 \\
            prefect $\to$ orchest & -1.04 \\
            \hline
        \end{tabular}
    \end{minipage}
\end{table}

\subsection{Results of Experiment}

For each 3! permutation, 15,756 activity series estimated 5,252 SVAR models, considering period divisions and shifts. As previously discussed, the final number of groups was 187, with 87 in the REV group and 100 in the non-REV group. Consequently, $187 \times 3!$ score tables corresponding to each group were generated. Note that the ADF test and VAR estimation are invariant to variable ordering.

Table~\ref{tab:results-of-model-fitting} summarizes the distribution of parameters and various test statistics determined during the execution of MIAO Phase 1, grouped into three categories: time series stationarization, VAR estimation, and serial correlation by the Ljung–Box test.

The time series stationarization results show that ADF tests (n=15,756) identified over 90\% of series as stationary ($p<0.05$). The remaining unit root processes were converted to stationary processes using fractional differencing (N-Frac diff in the table). While most samples are stationary processes with $n = 0$, the remaining few percent of the unit root processes undergo fractional differencing with $n$ up to 0.8.

In the VAR results, the sample size indicates activity data length, ranging from 365 to 1,461 days due to the $m=4$ setting. The Lag represents the optimal lag order for each VAR, determined by the VAR\_BEST\_LAG function in MIAO Phase 1. 
We observed 5,252 samples of lags orders (Lag), selected based on the AIC, with a minimum value (Min) of 1, an average (Mean) of 10.93, and a maximum value (Max) of 15. The $n$\% denotes the value at that percentile, indicating that $n$\% of the data falls below this value and $(100-n)$\% falls above it. For instance, 25\% = 7.83 indicates that 25\% of the data points are less than or equal to 7.83, while 75\% exceed this value. The AIC, Bayesian information criterion (BIC), and Hannan–Quinn information criterion (HQIC) are the information criteria corresponding to the lag obtained from the VAR\_BEST\_LAG function. Given that there is no extreme difference in the distributions of the three ICs and that 98\% of the cases selected the same lag across all criteria, the lag selection is considered robust.

The maximum lag order in the Ljung–Box test results is distributed between 10 and 30, as per the setting. The larger the test statistic, the stronger the evidence to reject the null hypothesis. With a set significance level of 10\%, we conclude that no serial correlation exists in the VAR residuals when $p \geq 0.1$. In the table, the significance level is exceeded at the 10\% point, indicating that for $90+\alpha\%$ of the data, the reliability of the VAR estimation accuracy is high.

Table~\ref{tab:results-of-miao} presents examples of score tables obtained from MIAO. In Group 1, if PyTorch is designated as $C_2$, then $\mathrm{MS}_{C_2T} > 0 \land \mathrm{MS}_{TC_2} < 0$, while all other scores are negative. This indicates a substantial negative impact from $C_2$ to $T$. In contrast, all scores for Group 24 are negative, showing that all the OSS projects are exerting a mutually positive influence on each other.

\subsection{RQ1: How accurately can MIAO identify OSS projects that ceased development due to competitive influences?}

\begin{table*}[t]
    \centering
    \caption{Summary of classification metrics across $3!$ sets}
    \begin{tabular}{|l|l|rrrrrr|r|r|}
    \hline
     \textbf{EVAL No.} & \textbf{Indicator} & \textbf{Set 1} & \textbf{Set 2} & \textbf{Set 3} & \textbf{Set 4} & \textbf{Set 5} & \textbf{Set 6} & \textbf{Mean$\pm$SD} & \textbf{Support} \\ \hline

    \multirow{7}{*}{EVAL1: Retrospective} 
        & Accuracy & \rvz{0.67} & \rvz{0.68} & \rvz{0.74} & \rvz{0.74} & \rvz{0.78} & \rvz{0.78} & \rvz{0.73 $\pm$ 0.05} & 187.0 \\
        & REV F1-Score & \rvz{0.67} & \rvz{0.62} & \rvz{0.73} & \rvz{0.72} & \rvz{0.75} & \rvz{0.74} & 0.71 \rvz{$\pm$ 0.05} & 87.0 \\
        & REV precision & \rvz{0.62} & \rvz{0.70} & \rvz{0.70} & \rvz{0.72} & \rvz{0.80} & \rvz{0.80} & \rvz{0.72 $\pm$ 0.07} & 87.0 \\
        & REV recall & \rvz{0.74} & \rvz{0.56} & \rvz{0.76} & \rvz{0.71} & \rvz{0.70} & \rvz{0.69} & \rvz{0.69 $\pm$ 0.07} & 87.0 \\
        & non-REV F1-Score & \rvz{0.66} & \rvz{0.73} & \rvz{0.75} & \rvz{0.76} & \rvz{0.81} & \rvz{0.80} & \rvz{0.75 $\pm$ 0.05} & 100.0 \\
        & non-REV precision & \rvz{0.73} & \rvz{0.68} & \rvz{0.77} & \rvz{0.75} & \rvz{0.77} & \rvz{0.76} & 0.74 $\pm$ \rvz{0.04} & 100.0 \\
        & non-REV recall & \rvz{0.61} & \rvz{0.79} & \rvz{0.72} & \rvz{0.76} & \rvz{0.85} & \rvz{0.85} & \rvz{0.76 $\pm$ 0.09} & 100.0 \\
    \hline

    \multirow{7}{*}{EVAL2: Predictive} 
        & Accuracy & \rvz{0.67} & \rvz{0.74} & \rvz{0.70} & \rvz{0.64} & \rvz{0.58} & \rvz{0.74} & \rvz{0.68 $\pm$ 0.06} & 159.0 \\
        & REV F1-Score & \rvz{0.56} & \rvz{0.65} & \rvz{0.60} & \rvz{0.48} & \rvz{0.54} & \rvz{0.64} & \rvz{0.58 $\pm$ 0.07} & 59.0 \\
        & REV precision & \rvz{0.55} & \rvz{0.66} & \rvz{0.60} & \rvz{0.52} & \rvz{0.46} & \rvz{0.66} & \rvz{0.57 $\pm$ 0.08} & 59.0 \\
        & REV recall & \rvz{0.58} & 0.64 & 0.59 & \rvz{0.44} & \rvz{0.64} & 0.63 & \rvz{0.59 $\pm$ 0.08} & 59.0 \\
        & non-REV F1-Score & \rvz{0.73} & \rvz{0.80} & \rvz{0.77} & \rvz{0.73} & \rvz{0.62} & \rvz{0.80} & \rvz{0.74 $\pm$ 0.06} & 100.0 \\
        & non-REV precision & \rvz{0.74} & \rvz{0.79} & \rvz{0.76} & \rvz{0.70} & \rvz{0.72} & 0.79 & \rvz{0.75 $\pm$ 0.04} & 100.0 \\
        & non-REV recall & \rvz{0.72} & \rvz{0.80} & \rvz{0.77} & \rvz{0.76} & \rvz{0.55} & \rvz{0.81} & \rvz{0.73 $\pm$ 0.10} & 100.0 \\
    \hline
    \end{tabular}
    \label{tab:classification_report}
\end{table*}

To answer RQ1, we evaluated the performance of MIAO from multiple angles using the following two evaluations.

\textbf{EVAL1 (Full Data): Evaluating the classification performance for distinguishing between REV and non-REV.} We constructed a decision tree model for REV classification using the MIAO score tables as explanatory variables over the whole time frame of the projects.

\textbf{EVAL2 (Masked Data): Evaluating predictive performance by masking the final year of data.}
We construct a dataset by masking one year from the End date in Table~\ref{tab:dataset} and run the same experiment as in Section~\ref{sec:experiment} to obtain a new set of MIAO score tables. Using these new scores as explanatory variables, we solve the same classification problem as in EVAL1 to assess how accurately MIAO can predict future REV from a point one year prior. If the model's classification performance is still strong at this earlier point, it suggests that MIAO may be able to detect the early signs of REV.

We now turn to the preparations for EVAL1, explaining how variables for the decision tree were prepared and how groups spanning divided periods were handled. The objective variable for the decision tree is the REV column in Table \ref{tab:dataset}, and the model predicts whether a target group has REV or non-REV. using $187 \times 6$ $\mathrm{AMS}_{ij}$ values that obtained from experiment in Section \ref{sec:experiment}. However, as variable names change for each score table, we unify them as $c1 \to t$, $c2 \to t$, $t \to c1$, $t \to c2$, $c1 \to c2$, and $c2 \to c1$. Here, $t$ represents the target, while $c1$ and $c2$ represent competitors 1 and 2, respectively.

The generated MIAO score table consists of group-specific values and is not normalized to allow comparisons across groups. To address this, we 1) normalize SCE, 2) integrate across \#splits, and 3) mitigate bias arising from the length of $|T_m|$. For 1), SCE is the cumulative response of $i$ to a unit structural shock in $j$. Therefore, by dividing $\mathrm{SCE}_{ij}$ in Equation~\ref{eq:miao-score} by the standard deviation $\sigma_i$ of $i$, we obtain a unit-free scale indicating how many multiples of its own standard deviation $i$ moves in response to a shock in $j$. For 2), we integrate into a single score table by summing all the score tables obtained for each \#split. For 3), we mitigate bias due to differences in the length of $m$—that is, the number of times scores are summed—by normalizing AMS by dividing it by $m$. For groups where \#split integration occurs, we divide the integrated AMS by the total of $m$ across all \#splits. The final AMS obtained through these processes is denoted as $\tilde{\mathrm{AMS}}$, ensuring consistency of the score scale across all groups through it.

The learning and evaluation of the classification model were performed using cross-validation to improve the generalization performance. 
Owing to the sample size of 187 is not large enough, we used leave-one-out cross-validation (LOOCV) \cite{10.5555/1643031.1643047}. 
Therefore, the various indicators shown in this section are average values of the indicators obtained in each iteration. 

Table \ref{tab:classification_report} presents the model’s average performance across all 3! permutations for EVAL1 and EVAL2. Each Set \textit{N} corresponds to one permutation and reports the mean accuracy and the class‑wise mean F1‑scores for that set. 
In EVAL2, masking the most recent year reduced the number of effective samples to 59. Since both evaluation sets have a class imbalance, we applied weights to the samples using the formula $w_c = n_{samples}/(n_{classes}\cdot n_c)$ to mitigate the effect of the imbalance. 

Across all $3!$ permutations in EVAL1, the model achieved an average accuracy of \rvz{0.73} (range: 0.67--\rvz{0.78}). The F1 scores averaged 0.71 for the REV class (range: \rvz{0.62--0.75}) and \rvz{0.75} for the non-REV class (range: \rvz{0.66--0.81}). Such robust performance demonstrates the viability of our novel methodology for retrospectively identifying REV cases. Based on the REV F1-Score, the best performance was observed in Set \rvz{5}, while the worst was in Set \rvz{2}. Let us denote the competitor with more commits as \(c^+\) and the one with fewer as \(c^-\). The variable orderings for these sets are: Set \rvz{5} \rvz{(\(t, c^-, c^+\))} and Set \rvz{2} \rvz{(\(c^-, c^+, t\))}. A key difference is that in the lower-performing \rvz{set, both competitors precede the target}, \rvz{whereas the higher-performing set places the target first}. This suggests that arranging variables such that \rvz{both competitors precede the target} may make it more difficult to distinguish the REV class.

In EVAL2, the model achieved an average accuracy of \rvz{0.68} (range: \rvz{0.58--0.74}). The F1 scores averaged \rvz{0.58} for the REV class (range: \rvz{0.48--0.65}) and \rvz{0.74} for the non-REV class (range: \rvz{0.62--0.80}). The predictive performance showed greater variability, indicating that the variable ordering became even more critical. However, the accuracy of \rvz{0.74} in \rvz{Set 2 and Set 6} is a \rvz{moderate} result, suggesting that MIAO can be used as an automated screening tool. This \rvz{predictive} score, even with data from one year prior, indicates that MIAO can serve as \rvz{part of} an effective early-warning system. By applying it for continuous, periodic observation, the detection accuracy of an impending REV could be progressively enhanced. Conversely, \rvz{Set 4} exhibited the lowest performance in the F1-Score for the REV class (\rvz{0.48}). 

\begin{shadedframe}
\noindent Answer to RQ1: MIAO accurately identifies REV, achieving up to \rvz{0.78} accuracy in retrospective classification and \rvz{0.74} accuracy in predictive tasks one year prior.
\end{shadedframe}

\subsection{RQ2: What are the key factors of competitive relationships that affect the survival of OSS projects?}

\begin{figure*}[t]
\centering
\begin{minipage}[b]{0.53\textwidth}
    \centering
    \includegraphics[width=\linewidth]{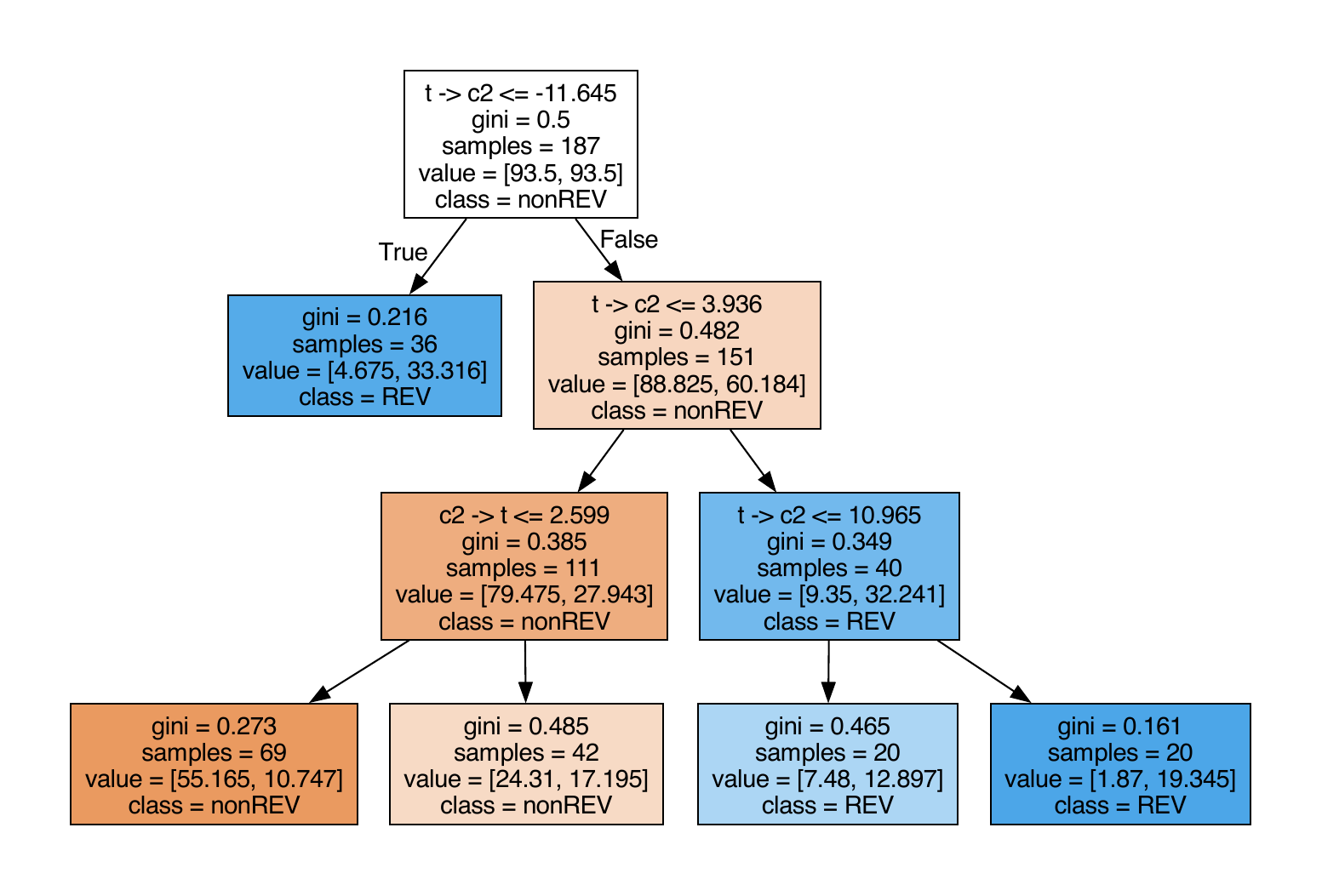}
    \subcaption{EVAL1, Set \rvz{5}}
    \label{fig:decision-tree-eval1}
\end{minipage}
\hfill
\begin{minipage}[b]{0.38\textwidth}
    \centering
    \includegraphics[width=\linewidth]{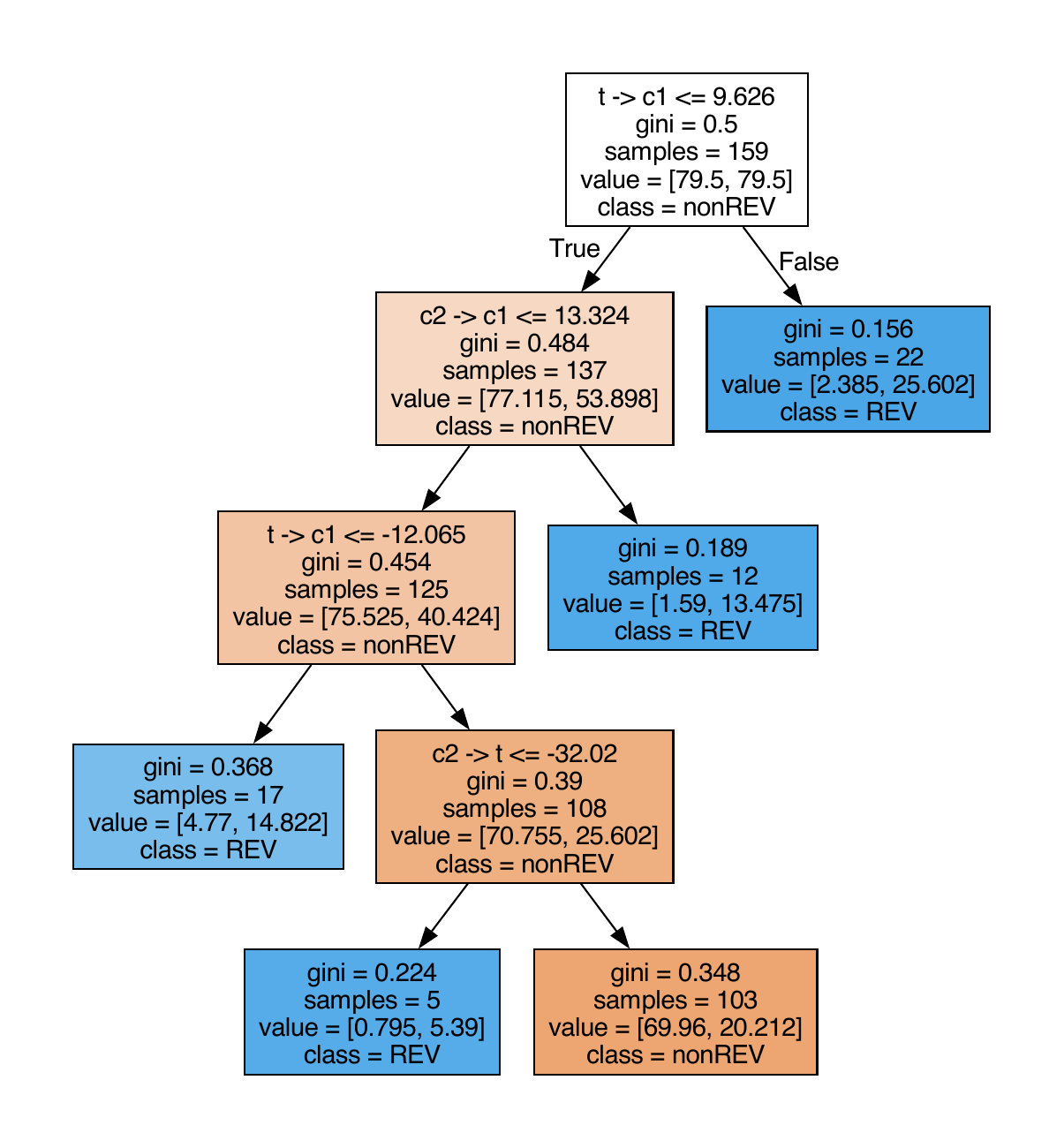}
    \subcaption{EVAL2, Set \rvz{2}}
    \label{fig:decision-tree-eval2}
\end{minipage}
\caption{Decision Trees of the Best-Performing Set}
\label{fig:decision-trees}
\end{figure*}

\vspace{0.4em}

\begin{table}[t]
    \centering
    \caption{Feature importances of The Best-performing Sets }
    \label{tab:feature-importance}
    \begin{minipage}{0.53\columnwidth}
        \centering
        \begin{tabular}{|c|c|} \hline
            \textbf{EVAL1, Set \rvz{5}} & \textbf{Contribution} \\ \hline
            $t \to c2$ & $\rvz{0.91}$ \\
            \rvz{$c2 \to t$} & $\rvz{0.09}$ \\
            \hline
        \end{tabular}
    \end{minipage}%
    \hfill
    \begin{minipage}{0.47\columnwidth}
        \centering
        \begin{tabular}{|c|c|} \hline
            \textbf{EVAL2, Set \rvz{2}} & \textbf{Contribution} \\ \hline
            \rvz{$t \to c1$} & $\rvz{0.60}$ \\
            \rvz{$c2 \to c1$} & $\rvz{0.25}$ \\
            \rvz{$c2 \to t$} & $\rvz{0.15}$ \\
            \hline
        \end{tabular}
    \end{minipage}
\end{table}

Our second research question explores the key factors of competitive relationships that affect the survival of OSS projects. To answer this question, we analyzed the decision trees for EVAL1 and EVAL2. Since the $AMS$ is normalized by $\sigma_i$ to enable cross-group comparisons, the thresholds in these decision trees are not specific to this experiment but can be considered generally applicable. In other words, the threshold at each node itself is a significant value for discriminating between REV and non-REV. We can answer RQ2 by systematizing the ranges or thresholds that the MIAO scores fall.

Figure~\ref{fig:decision-trees} visualizes the decision trees for the best-performing sets in EVAL1 and EVAL2, respectively, and Table~\ref{tab:feature-importance} shows the corresponding feature importance.
Looking at the EVAL1 decision tree (Figure~\ref{fig:decision-tree-eval1}), the conditions for identifying REV are characterized by the target's influence on the competitor. The probability of REV increases when \rvz{$(t \to c2 \leq -11.645) \lor (t \to c2 > 3.936)$}. In essence, \rvz{REV cases are concentrated in both tails of the $t \to c2$ distribution, while non-REV cases tend to cluster in the middle range between the two decision-tree thresholds.} The feature importance of EVAL1 (Table~\ref{tab:feature-importance}) confirms this, showing that the differences between classes can be explained almost entirely by $t \to c2$\rvz{, with a smaller contribution from $c2 \to t$}. This result suggests, albeit counter-intuitively, that the commonly assumed negative impact of the competitor on the target did not contribute \rvz{substantially} in this retrospective analysis, but actually the other way around.
As concrete examples of true positives, Group 9 ($t$: Sick-Beard, $c2$: Sonarr) with $t \to c2 = \rvz{-71.18}$ falls into the range \rvz{$(t \to c2 \leq -11.645)$}, while \rvz{Group 2 ($t$: Voldemort, $c2$: Venice) with $t \to c2 = 20.69$} falls into \rvz{$(t \to c2 > 3.936)$}.

Next, using the EVAL2 decision tree (Figure~\ref{fig:decision-tree-eval2}), we analyze the differences from EVAL1. The results for EVAL2 are \rvz{less concentrated than in EVAL1: the target-to-competitor feature $t \to c1$ serves as the primary distinguishing feature for REV, but other relationships also contribute. Specifically, the largest REV-class leaves occur when $(t \to c1 > 9.626)$ (accounting for 32\% of the REV class), when $(t \to c1 \leq -12.065 \land c2 \to c1 \leq 13.324)$ (accounting for 19\%), and when $(t \to c1 \leq 9.626 \land c2 \to c1 > 13.324)$ (accounting for 17\%). A smaller REV-class leaf also involves $c2 \to t$ (accounting for 7\%). These paths indicate that competitor-to-competitor and competitor-to-target relationships also contribute. The feature importance of EVAL2 (Table~\ref{tab:feature-importance}) reflects this trend, showing that $t \to c1$ contributes most, while $c2 \to c1$ and $c2 \to t$ also contribute.}

\rvz{Taken together, the comparison between EVAL1 and EVAL2 suggests that REV is characterized by a shift in the structure of competitive influence. In EVAL2, although $t \to c1$ is the primary feature, the classification signal tends to extend to relationships from competitors to the target and between competitors, indicating that early warning signs of REV still appear as a mixed competitive structure. In contrast, EVAL1 is dominated by the extreme values of $t \to c2$, suggesting that, in the full observation period, the signal becomes concentrated on target-to-competitor influence. These results suggest that the key factor of competitive relationships that affects the survival of OSS projects is unidirectional target-to-competitor influence, and that this influence may become more pronounced over time.} This indicates that MIAO could \rvz{help capture the} process from the initial signs of REV to the subsequent decline after its occurrence.

\begin{shadedframe}
\noindent Answer to RQ2: The key factor of competitive relationships that affects the survival of OSS projects is the degree of unidirectional influence from the target to the competitor.
\end{shadedframe}

\section{Discussion and Implications}
Our proposed approach has the following implications for both software engineering researchers and software practitioners taking care of OSS projects.

First, for practitioners managing OSS projects, MIAO serves as a quantitative early-warning system. By monitoring external competitive pressures, maintainers can move beyond reactive measures and make proactive strategic decisions, such as adjusting their roadmap or forming alliances, to ensure their project's survival and relevance in a changing ecosystem.

Second, for practitioners adopting OSS, our method functions as an advanced risk assessment tool. It allows organizations to evaluate a project's long-term viability by analyzing its position within its competitive landscape. This helps select resilient technologies and avoid projects showing early signs of decline from competition.

Finally, for software engineering researchers, this work provides a novel, scalable methodology to quantitatively study OSS ecosystem dynamics. It enables a shift from analyzing internal project factors to modeling the external competitive relationships that drive software evolution, opening new research avenues into the rise and fall of technologies at an ecosystem level.

\section{Related Work}

In research on the influence of ecosystem-level external factors on OSS projects, Valiev et al. \cite{10.1145/3236024.3236062} showed that a project’s relative position within the package-dependency network affects its survival probability. Allaho et al. \cite{6785687} analyzed contributor social-network structures and found a positive correlation between the number of developers’ social ties and developer productivity. Zhao et al. \cite{Yuhang2021} evaluated the impact of sponsorship and discovered a positive correlation between project activity levels and sponsorship. Ait et al. \cite{9796216} conducted a survival analysis of 1,127 projects and found that more than half ceased activity within the first four years; however, projects backed by organizations (e.g., companies) or supported by larger communities exhibited statistically higher survival rates. However, these studies do not directly address competitive relationships between projects.

Indeed, Curto-Millet et al. \cite{Curto-Millet03092023} position competition for finite volunteer developers as a key example of the ``resource-based'' dimension of open-source sustainability, noting that when developer effort concentrates on one project, it can deprive others. This aligns with Coelho et al.’s \cite{Coelho_2017} finding that the emergence of strong competitors is a leading driver of abandonment. Nevertheless, their qualitative frameworks and analyses do not offer a means to empirically quantify such dynamics across large ecosystems. To address this gap, we introduce MIAO, which models (activity-)shock propagation between projects and thus provides a scalable, data-driven measure of competitive pressure across OSS communities. 

\section{Threats to Validity}

\textit{Internal Validity:~} 
The first threat is the accuracy of the dataset. The REV class targets projects with migration destinations in their README, but cannot definitively confirm actual REV decline. While the dataset likely contains genuine REV cases, some may be false positives. However, excluding small projects during filtering increases the likelihood of selecting projects with competitive experience. Additionally, manual curation data validity remains a concern for both REV and non-REV classes, as non-competitors may have been inadvertently included despite careful verification.
Using only daily commit counts as a proxy for project activity may not fully represent the liveness of the projects, and other signals (e.g., star trajectories) may also be considered. 
However, commits are a standard trace of OSS activity; Kolassa et al. \cite{10.1145/2491055.2491073} quantified commit-frequency distributions at scale and demonstrated their usefulness as fundamental statistics for understanding development processes. Furthermore, Decan et al. \cite{DECAN2020110573} used commit activity forecasts to assess developer abandonment risk, a key threat to project sustainability. 
Another threat is the confounding of competitive pressure and other factors. Although MIAO uncovers statistical associations for REV projects, it does not disentangle competitive decline from other mechanisms (e.g., technological obsolescence). A detailed causal analysis of REV cases is required for this, which is left for future work.

\textit{External Validity:~}
The analysis and the results are confined to 187 projects. Thus, the results may not hold for different time ranges or project populations of other sizes. Moreover, our findings are based on the open-source projects on GitHub and may not be generalized to commercial projects.
In addition, our scope intentionally excludes fork-based competition. Forking is a common mechanism of OSS evolution and competition; however, fork-based cases share an identical pre-fork history between the target and the forked competitor, yielding perfect multicollinearity that prevents unique identification of VAR coefficients. While one could analyze only post-fork behavior, many fork-based cases in our candidates did not satisfy our minimum temporal-overlap requirement after the fork. Therefore, our findings should be interpreted as characterizing \emph{independent} competitive relationships (non-fork competition), and may not directly generalize to competitive dynamics that primarily manifest through forks.
Finally, to ensure comparability across projects, we fixed the model dimensionality to a three-variable setting (one target and two competitors), where the two competitors were selected using our competitor-identification procedure. In some cases, this procedure relies on LLM assistance, which may miss relevant competitors or include imperfect matches, potentially attenuating estimated competitive effects. Moreover, fixing the dimensionality to two competitors may under-represent ecosystems in which multiple alternatives simultaneously attract activity and may not capture higher-order competitive interactions among many competitors. While higher-dimensional models are technically feasible, evaluating them in a comparable manner across projects with varying numbers of competitors is non-trivial.

\section{Conclusion and Future Work}

Although studies emphasize \emph{internal} determinants of OSS sustainability, our MIAO framework shows that \emph{external} competitive dynamics can be quantified to explain OSS decline. MIAO (Mutual Impact Analysis of OSS) is our proposed econometric method to quantify these competitive dynamics. We use this framework to identify projects that ceased development due to competition, which we define as Rising Events(REV). From mining GitHub across 187 project groups, our analysis reveals that MIAO reaches \rvz{78\%} accuracy in retrospective REV classification and \rvz{74\%} in one-year-ahead prediction, the degree of unidirectional influence from the target to the competitor emerging as the key factor. 

Building on this foundation, we see three immediate directions for extension. 
\emph{(i)~Dataset expansion:} enlarging the sample and applying a causal-inference-aware labelling protocol to separate competitive decline from other cessation factors will strengthen external and internal validity.
\emph{(ii)~Alternative activity signals:} incorporating issue traffic, release cadence, or dependency updates alongside commits  could model the interplay between these variables, potentially elucidating more complex REV mechanisms.
\emph{(iii)~Fork-based competition:} developing methods to resolve the perfect multicollinearity inherent in shared pre-fork histories will allow us to generalize our findings to dynamics driven by forks, a key evolutionary mechanism.

\textbf{Data availability statement}: Our entire dataset, experiment results, and accompanying scripts are publicly available \cite{our-artifact}.

\bibliographystyle{ACM-Reference-Format}
\bibliography{ref}

\end{document}